\documentstyle[12pt]{article}
\oddsidemargin--1mm
\begin{document}
\newcommand{\pl}{\partial}
\newcommand{\be}{\begin{equation}}
\newcommand{\ee}{\end{equation}}
\newcommand{\ba}{\begin{eqnarray}}
\newcommand{\ea}{\end{eqnarray}}
\newcommand{\mbf}[1]{\mbox{\boldmath$ #1$}}
\def\<{\langle}
\def\>{\rangle}

\begin{center}
{\Large Yang-Mills theory as an Abelian theory without gauge fixing}

\vskip 0.5cm
Sergei V. Shabanov {\footnote{on leave from Laboratory of
Theoretical Physics, JINR, Dubna, Russia}}

\vskip 0.2cm

{\em L.P.T.H.E., Universit\'e Pierre et Marie Curie, 4 place Jussieu,\\
Tour 16, 1er etage, Paris Cedex 05, F-75252, France}
\end{center}

\begin{abstract}
A general procedure to reveal an Abelian structure of Yang-Mills theories
by means of a (nonlocal) change of variables, rather than by gauge fixing,
in the space of connections is proposed.
The Abelian gauge group is isomorphic to the maximal Abelian
subgroup of the Yang-Mills gauge group, but not its subgroup. 
 A Maxwell field of the Abelian theory
contains topological degrees of freedom of original
Yang-Mills fields which generate monopole-like
and flux-like defects upon an Abelian projection.
 't Hooft's conjecture
that ``monopole'' dynamics is projection independent is
proved for a special class of Abelian projections.
A partial duality and a dynamical regime
in which the theory may have massive excitations being 
knot-like solitons are discussed.
\end{abstract}

{ \bf 1. General remarks}. 
One of the  physical scenarios of the color confinement 
is based on the idea that the vacuum state of quantum Yang-Mills theory  
is realized by a condensate of monopole-antimonopole pairs \cite{1}. In such
a vacuum the field between two colored sources would be squeezed into
a tube whose energy is proportional to its length. The picture is
dual to the magnetic monopole confinement in a superconductor of the
second kind. Monopoles as classical solutions with finite energy
are absent in a pure Yang-Mills theory. To realize the dual scenario of
the confinement, 't Hooft proposed an Abelian projection where the gauge
group is broken by a suitable gauge condition to its maximal Abelian
subgroup \cite{2}. Since the topology of the SU(N) manifold and that of its
maximal Abelian subgroup [U(1)]${}^{N-1}$ are different, any such gauge
is singular, meaning that a gauge group element which
transforms a generic SU(N) connection onto the gauge fixing surface
is not regular everywhere in spacetime. The singularities may form
worldlines that are usually interpreted as worldlines of magnetic monopoles
(whose charges are defined 
with respect to the unbroken Abelian subgroup). As a result
the original Yang-Mills theory turns into electrodynamics with magnetic
monopoles. Recent numerical simulations show that the monopole degrees
of freedom in the Abelian projection can indeed form a condensate
responsible for the confinement \cite{3}. 

Although the numerical results look rather appealing and stimulating, 
they still do not provide us with an  
understanding of the confinement mechanism and
a nonperturbative spectrum in Yang-Mills theory.
In particular, monopoles seem to 
emerge as an {\em artifact} of gauge fixing.
The Abelian group appears as a {\em subgroup} of the full Yang-Mills
gauge group. However one can easily construct colored states
which are singlets with respect to the unbroken (maximal) Abelian
subgroup, and, hence, they would not be confined even if the ``monopoles''
condense.
A choice of the gauge may be convenient in practical computations.
However, no physical phenomenon can depend on it.
This suggests that in Yang-Mills theory there seems to be a new mechanism
of confinement at work which has yet to be understood, and the
reason of why Abelian projections work so well in the lattice
theory must be explained in a gauge independent way.
A first and necessary step in this direction is to reveal
an Abelian structure of Yang-Mills theory without any gauge fixing.

In this letter, a Yang-Mills theory is reformulated as an Abelian
gauge theory via a (nonlocal) change of variables in
the space of connections, rather than via a gauge fixing (or
an Abelian projection). 
In particular, it turns out to be possible to construct
the field variables in
the Abelian theory so that they are invariant under the original
non-Abelian gauge transformations.  
Therefore an effective Abelian structure is inherent
to the Yang-Mills theory and gauge independent.
An Abelian vector potential carries some topological degrees
of freedom of the original Yang-Mills connection which
generate monopole-like and flux-like defects upon an 
Abelian projection (in which the gauge group is broken to its
maximal Abelian subgroup by a {\em gauge fixing} \cite{2}). 
For a rather wide class of Abelian projections, which have
a characteristic property that topological defects occur
in the Abelian components of projected connections,
we offer theoretical arguments to prove 't Hooft's conjecture
that dynamics of ``monopoles'' does not depend on the choice
of a projection, i.e., it is gauge independent.

A generalization 
of the parameterization of the Yang-Mills connection proposed by Faddeev and
Niemi \cite{5} is considered as a special example. While
revealing a partial duality in Yang-Mills theory, 
it has an important advantage that it is a {\em genuine}
change of variables in the functional integral. Therefore it provides
a description of the {\em off-shell} dynamics
of physical degrees of freedom which is compatible with
the Gauss law.
Following the Wilsonian arguments of \cite{5},
we discuss the partial duality in the theory and 
a dynamical regime in which the topological degrees of
freedom may form massive excitations being knot-like solitons.

{\bf 2. Gauge group SU(2)}. 
Let $\mbf{ A}_\mu$
be an SU(2) connection. Consider the following parameterization
of the connection
\be
\mbf{ A}_\mu = \mbf{\alpha}_\mu + \mbf{ n}C_\mu +
 \mbf{W}_{\mu}\ ,
\ \ \ \ \mbf{\alpha}_\mu =g^{-1}\pl_\mu \mbf{ n}\times \mbf{ n}\ ,\ \ \ \
\mbf{W}_\mu \cdot \mbf{n} =0\ ,
\label{1}
\ee
where $g$ is a coupling constant, 
$\mbf{\alpha}_\mu$ is a connection introduced by Cho \cite{4}, $\mbf{ n}$
is a unit isotopic vector, $\mbf{ n}^2=1$. The dot and cross stand,
respectively, for the dot and cross products in the isotopic space
whose elements are denoted by boldface letters. 
Relation (\ref{1}) is
not yet a genuine {\em change} of variables in the affine space
of connections. Two more conditions have to be imposed on $\mbf{W}_\mu$
in order for (\ref{1}) to be a change of variables. We may set in general
\be 
\mbf{\chi}(\mbf{W}, \mbf{n}, C)=0\ , \ \ \ \ 
\mbf{\chi}\cdot\mbf{n}\equiv 0\ .
\label{1a}
\ee
The function $\mbf{\chi}$ can be chosen so that a
solution of Eq. (\ref{1a}) determines a {\em local} and {\em explicit}
parameterization of 
eight components in $\mbf{W}_\mu$ by
six functional variables (see section 5), 
thus leading to a generalization of the
parameterization given in \cite{5}.
We will also show that
some $\mbf{\chi}$'s, for which Eq. (\ref{1a}) admits 
{\em nonlocal} parameterizations
of the SU(2) connection, can naturally be associated with 't Hooft's
Abelian projections where an Abelian vector potential contains
magnetic monopoles described by the field $\mbf{n}$.

Before specifying $\mbf{\chi}$ 
let us first analyze the gauge transformation law of the new variables.
An infinitesimal  gauge transformation 
of the SU(2) connection reads
\be
\delta \mbf{ A}_\mu = g^{-1}\nabla_\mu (\mbf{ A})\mbf{\varphi}=
g^{-1}\left[\pl_\mu \mbf{\varphi} + g\mbf{ A}_\mu \times \mbf{\varphi}
\right]\ .
\label{3}
\ee
From (\ref{1}) we infer
\be 
C_\mu = \mbf{n}\cdot \mbf{A}_\mu\  ,\ \ \ \
\mbf{W}_\mu = g^{-1}\mbf{n}\times \nabla(\mbf{A})\mbf{n}\ .
\label{3a}
\ee
Substituting these relations into (\ref{1a}) and solving them for 
$\mbf{n}$ (two equations for two independent variables in $\mbf{n}$),
we find $\mbf{n}=\mbf{n}(\mbf{A})$. The latter together with (\ref{3a})
specifies the inverse change of variables. Let $\delta\mbf{n}$ be
an infinitesimal gauge transformation of $\mbf{n}$. Then from 
(\ref{3a}) and (\ref{3}) it follows that
\ba
\delta C_\mu &=& \mbf{A}_\mu\cdot(\delta\mbf{n}-\mbf{n}\times\mbf{\varphi})
+ g^{-1}\mbf{n}\cdot \pl_\mu\mbf{\varphi}\ ,\label{4}\\
\delta\mbf{W}_\mu &=&
\mbf{W}\times\mbf{\varphi} - \mbf{n}[\mbf{W}_\mu\cdot(\delta\mbf{n}-
\mbf{n}\times\mbf{\varphi})] + g^{-1}\mbf{n}\times \pl_\mu(
\delta\mbf{n}-\mbf{n}\times\mbf{\varphi})\ ,\label{5}
\ea
where we have used that $\mbf{n}\cdot\delta\mbf{n}=0$.
An explicit form of $\delta\mbf{n}$ can be found 
from the equation $\delta\mbf{\chi}(\mbf{n},\mbf{A})=0$
(taken on the surface $\mbf{\chi}=0$)
where $\mbf{\chi}(\mbf{n},\mbf{A})$ is obtained by a substitution
of (\ref{3a}) into $\mbf{\chi}(\mbf{W},C,\mbf{n})$. 
We emphasize that $\delta\mbf{n}$ is determined by the choice of
$\mbf{\chi}$ and so are $\delta C_\mu$ and $\delta
\mbf{W}_\mu$.

Let us introduce
a local orthonormal basis in the isotopic space 
$\mbf{n}$, $\mbf{ n}_r$ and $\mbf{ n}_r^*$:
$\mbf{n}_r\cdot \mbf{n}=0$, $\mbf{n}_r^2=0$ and 
$\mbf{n}_r\cdot\mbf{n}_r^*=1$.
We also have $\mbf{n}\times\mbf{n}_r =i\mbf{n}_r$ and $\mbf{n}_r\times
\mbf{n}_r^*=i\mbf{n}$. With $\mbf{n}$ fixed, the basis is determined
modulo {\em local} transformations
\be 
\mbf{n}_r \rightarrow e^{i\xi}\mbf{n}_r\ .
\label{5a}
\ee 
It should be noted that this gauge freedom is {\em not}
associated with the gauge group of the Yang-Mills theory
because the new variables remain unchanged under (\ref{5a}).
The local basis may not exist globally and the field
$\mbf{n}_r$ may have singularities. The reason is as
follows. At the spatial infinity, the connection
must be a pure gauge. Therefore $\mbf{n}$ becomes a
constant as $\vert\mbf{x}\vert $ approaches infinity, say,
$\mbf{n}_0=(0,0,1)$. The field $\mbf{n}$ is a map of
the three-sphere $S^3$, being the compactified space,
to the target two-sphere $S^2$ in the isotopic space.
The homotopy group $\pi_3(S^2)\sim Z$ is not trivial.
Integers from $Z$ are given by the Hopf invariant.
If one attempts to transform $\mbf{n}$ to $\mbf{n}_0$
everywhere in space by rotation, the rotation matrix
will be ill-defined on some closed and, in general, knotted
contours. Another type of singularities is associated
with the homotopy group $\pi_2(S^2)\sim Z$ when the field $\mbf{n}$
is restricted on some $S^2$ being a subspace of $S^3$.
We will show that in the new variables (\ref{1}), the
Yang-Mills theory looks like an Abelian theory 
in which a Maxwell potential contains magnetic 
monopoles and fluxes associated with the nontriviality of $\pi_2(S^2)$
and $\pi_3(S^2)$, respectively. 

Consider the decomposition
\be
\mbf{W}_\mu = W_\mu^* \mbf{n}_r + W_\mu \mbf{n}_r^* \ .
\label{6a}
\ee
The fields strength is, by definition, $
\mbf{F}_{\mu\nu}(\mbf{A})= \pl_\mu\mbf{A}_\nu -\pl_\nu\mbf{A}_\mu
+g\mbf{A}_\mu\times\mbf{A}_\nu$. 
In particular, 
\ba
\mbf{F}_{\mu\nu}(\mbf{\alpha}+\mbf{n}C)&=& \mbf{n}(C_{\mu\nu}-H_{\mu\nu})\ ,
\ \ \ \ C_{\mu\nu}=\pl_\mu C_\nu -\pl_\nu C_\mu\ ,
\label{7a}\\
H_{\mu\nu}&=& g^{-1} \mbf{ n}\cdot(\pl_\mu\mbf{ n}\times\pl_\nu\mbf{ n})=
\pl_\mu H_\nu -\pl_\nu H_\mu- H_{\mu\nu}^{(st)}\ ,
\label{7}
\ea
where $H_\mu = ig^{-1}\mbf{ n}_r^*\cdot \pl_\mu \mbf{ n}_r=
H_\mu^*$ and
$H_{\mu\nu}^{(st)}= ig^{-1}\mbf{ n}_r^*\cdot [\pl_\mu,\pl_\nu]
\mbf{ n}_r=H_{\mu\nu}^{(st)*}$ which is the field strength of Dirac strings
associated with the singularities of the local basis.
For example, the Wu-Yang monopole configuration is determined by
$\mbf{ n}=\mbf{ x}/r$, $r^2=\mbf{ x}^2$ and $C_\mu =W_{\mu}=0$.
The Dirac string is extended along the negative part of the
$z$-axis (if $\mbf{ x}= (x,y,z)$). It is also not difficult
to give an example of $\mbf{n}$ for which Dirac strings would
form closed linked and/or knotted contours (see, e.g., \cite{knot}). 
Thus, the vector potential $H_\mu$ describes possible monopole-like
and closed-flux-like degrees of freedom to which we refer as to topological
degrees of freedom in the Yang-Mills theory.
After a modest computation
we obtain
\be
\mbf{F}_{\mu\nu}^2 = \left[C_{\mu\nu}-H_{\mu\nu} + ig(W_\mu^*W_\nu-
W_\nu^*W_\mu)\right]^2 + \left \vert D_\mu W_\nu- D_\nu W_\mu
\right\vert^2\ ,
\label{9}
\ee
where $D_\mu W_\nu =\pl_\mu W_\nu - ig A_\mu W_\nu$ is the U(1)
covariant derivative, $A_\mu=C_\mu -H_\mu$. 
The Abelian gauge transformations
have the form
\be
A_\mu \rightarrow A_\mu + g^{-1} \pl_\mu \xi \ ,\ \ \ \
W_\mu= e^{i\xi} W_\mu \ .
\label{11a}
\ee
The transformations (\ref{11a}) can obviously be generated by (\ref{5a}), and
therefore they are not from the original SU(2) gauge group. In contrast
to the Abelian gauge transformations (\ref{11a}), the SU(2)
transformations depend on a concrete parameterization
of $\mbf{W}_\mu$. 
Because of topological defects associated with the nontriviality
of $\pi_2(S^2)$, the Bianchi identity for the Abelian strength tensor
$F_{\mu\nu}= C_{\mu\nu}-H_{\mu\nu}$ is violated. Let
${}^\star\! F_{\mu\nu}$
be the dual tensor. Then one can define a conservative current
\be
J_\mu = \pl_\nu\, {}^\star\! F_{\mu\nu}\ ,\ \ \ \ \pl_\mu J_\mu =0\ .
\label{11b}
\ee
The conservation of the topological current $J_\mu$ indicates
the existence of the U(1) (magnetic) symmetry on the classical
level in the theory (\ref{9}). 

{\bf 3. Abelian projections}. An Abelian projection
is introduced by  imposing a gauge condition on $\mbf{A}_\mu$ that
breaks the gauge group SU(2) to its (maximal) Abelian subgroup U(1).
As has already been pointed out in section 1, a gauge group
element which transforms a generic connection to the gauge 
fixing surface in the space of connections is not regular everywhere
in spacetime. The transformed (or projected) connections
contain topological defects (singularities). 
Consider a special class of Abelian projections with the characteristic
property that
topological defects appear only in the Abelian components of the projected
potentials. Let us describe this class in the new variables (\ref{1}).
Had the gauge transformations of the field $\mbf{n}$ been
just isotopic rotations,
\be
\delta \mbf{n}=\mbf{n}\times \mbf{\varphi}\ ,
\label{12a}
\ee
then the connection $\mbf{\alpha}_\mu +\mbf{n}C_\mu$ would become
purely Abelian upon the projection $\mbf{n}\rightarrow \mbf{n}_0$
for {\em any} choice of $\mbf{\chi}$ which is {\em compatible}
with (\ref{12a}). Recall that in our formulation the gauge
transformation law for $\mbf{n}$ depends on $\mbf{\chi}$.
The addition to the Abelian component $C_\mu$ resulting from
$\mbf{\alpha}_\mu$ upon the projection 
determines exactly the same topological 
defects as the connection $H_\mu$ in the Abelian theory (\ref{9}),
i.e., $\mbf{\alpha}_\mu + \mbf{n} C_\mu\rightarrow \mbf{n}_0 A_\mu$.

Now we show that for {\em every} Abelian projection from
the special class defined above one can construct
$\mbf{\chi}$ in (\ref{1a}) which determines a special parameterization of
$\mbf{W}_\mu$ such that (\ref{12a}) holds. Moreover the topological 
current
(\ref{11b}) is {\em invariant} under the SU(2) gauge transformations
(\ref{4}), (\ref{5}) and (\ref{12a}). Thus, 
the very existence of the (magnetic) 
symmetry (\ref{11b}) is not at all related to any gauge fixing
in the theory. We first give an example
of $\mbf{\chi}$ associated with the so called maximal Abelian
projection \cite{map} which is mostly used in numerical studies of the 
``monopole'' dynamics:
\be
\mbf{\chi}=\nabla_\mu(\mbf{\alpha}+\mbf{n}C)\mbf{W}_\mu\ .
\label{12b}
\ee
The compatibility of (\ref{12b}) with (\ref{12a}) 
follows from the fact \cite{7}
that under the transformations (\ref{4}), (\ref{5}) and (\ref{12a})
the isovector (\ref{12b}) is covariant, $\delta\mbf{\chi}
= \mbf{\chi}\times \mbf{\varphi}$, and it also fulfills the 
condition $\mbf{\chi}\cdot \mbf{n}\equiv 0$ as one can easily be
convinced by a direct computation. Upon the projection
$\mbf{n}\rightarrow \mbf{n}_0$, $\mbf{\chi}$ turns into
the maximal Abelian gauge condition. The Abelian part
of the connection equals $A_\mu$ and contains magnetic monopoles whose
charges are defined with respect to the unbroken U(1) subgroup
(rotations about $\mbf{n}_0$) \cite{7}. By construction, 
the corresponding conservative 
monopole current coincides with (\ref{11b}).

Suppose an Abelian projection is specified by
a gauge condition $\mbf{\chi}(\mbf{n}_0A^{(0)},\mbf{W}^{(0)})=0$,
where $\mbf{A}_\mu =\mbf{n}_0 A_\mu^{(0)} + \mbf{W}_\mu^{(0)}$
and $\mbf{n}_0\cdot\mbf{W}_\mu^{(0)}=0$. We also assume
that $\mbf{\chi}$ is covariant (or even invariant) under
the Abelian gauge transformations $\delta_a A_\mu^{(0)}=
g^{-1}\pl_\mu\varphi$ and $\delta_a\mbf{W}_\mu^{(0)}=
\varphi \mbf{W}_\mu^{(0)}\times \mbf{n}_0$. This 
ensures that the gauge symmetry is broken to U(1).
Consider the change of variables
(\ref{1}) in which the condition (\ref{1a}) is obtained
 by a simple replacement $\mbf{n}_0A_\mu^{(0)}
\rightarrow \mbf{\alpha}_\mu + \mbf{n}C_\mu$ and
$\mbf{W}_\mu^{(0)}\rightarrow \mbf{W}_\mu$ in the above
gauge condition. By construction, 
the gauge transformation law (\ref{12a}) is guaranteed.
All topological degrees of freedom of the Yang-Mills theory,
which are singled out as magnetic monopoles upon the Abelian
projection, are contained in the Abelian vector potential 
$A_\mu$ of the Maxwell theory (\ref{9}). 

Thus, in the new 
variables the aforementioned special class of Abelian
projections is described by a single ``projection'' 
$\mbf{n}\rightarrow\mbf{n}_0$.

{}From (\ref{7a}) and (\ref{7}) it follows that 
\be
\delta J_\mu = \pl_\nu\,{}^\star\!\delta F_{\mu\nu} =0\ ,
\label{12c}
\ee
that is, the topological current (\ref{11b}) is {\em invariant}
under the SU(2) gauge transformations. Note that $C_\mu$
is not transformed by a simple gradient shift. The contribution
of non-Abelian gauge transformations to $\delta C_{\mu\nu}$
is compensated in $\delta F_{\mu\nu}$ by 
$\delta H_{\mu\nu}$ which is generated
by gauge rotations of the local basis
$\delta \mbf{n}_r =
\mbf{n}_r\times \mbf{\varphi}$, $\mbf{\varphi}\cdot\mbf{n}=0$.
The restriction on $\mbf{\varphi}$ has been imposed because
$\delta\mbf{n}=0$
if $\mbf{\varphi}=\mbf{n}\varphi$, while
$\delta C_\mu = g^{-1}\pl_\mu \varphi$. 
Therefore there are two
groups U(1) in the theory. One is associated with the subgroup of the gauge
group which preserve $C_{\mu\nu}$, 
$\delta C_{\mu\nu}=0$, while the other is given by
transformations (\ref{11a}). The Abelian potential
$A_\mu$ is {\em invariant} under the U(1) subgroup of
U(1)$\times$U(1) which is selected by the condition 
$\varphi = \xi$. The charged field $W_\mu$ is also
{\em invariant} under this U(1) subgroup. 
According (\ref{5}) and (\ref{6a}), the SU(2)
gauge transformations can be regarded as local
generic rotations of any {\em rigid} local basis in the isotopic
space, $\delta \mbf{n}_r = \mbf{n}_r\times \mbf{\varphi}$
(no restriction on $\mbf{\varphi}$), provided the $\mbf{\chi}$ in
(\ref{1a}) is such that (\ref{12a}) holds. 
In this case
all field variables in the Maxwell theory (\ref{9})
are invariant under the SU(2) Yang-Mills gauge group.     

{\bf 4. 't Hooft's conjecture}.
In lattice simulations one is interested in an effective 
dynamics of the topological degrees of freedom, i.e.,
in an effective theory of the field $\mbf{n}$ in our formulation.
Recently the dual scenario of the color confinement
in the lattice Yang-Mills theory  has 
been reported to occur in several Abelian projection \cite{dig}. 
All the projections studied have a characteristic property that 
monopole-like topological defects 
are contained in Abelian components of projected connections. 
This certainly supports 't Hooft's conjecture that 
all Abelian projections are equivalent \cite{2}. Can one find
theoretical arguments to prove this conjecture? 
Here we explain how the proof can be done.

In our parameterization all the Abelian projections
in question are described by 
one simple (singular) gauge condition $\mbf{n}=\mbf{n}_0$.
The difference between projections is related to
a {\em reparameterization} of $\mbf{W}_\mu$.
As we have a {\em genuine} change of variables in the functional
integral, we can, in principle, 
integrate out $\mbf{W}_\mu$, and get
an effective action for $\mbf{n}$ and $C_\mu$. From a technical point of view,
this procedure involves two important steps. First, one 
has to compute a Jacobian of the change of variables.
Second, a gauge has to be fixed, otherwise the integral is 
divergent. The latter can be done  by means of the conventional
Faddeev-Popov recipe with a nonsingular gauge (e.g. a background
or Lorentz gauge) before the change of variables.
The first problem is solved in the following
way \cite{7}. Consider the identity $1 =\int D\mbf{n}
\Delta(\mbf{A},\mbf{n})\delta(\mbf{\chi})$ where
$\mbf{\chi}=\mbf{\chi}(\mbf{A},\mbf{n})$ is obtained
by a substitution of (\ref{3a}) into $\mbf{\chi}(\mbf{W},C,\mbf{n})$.
Clearly, $\Delta(\mbf{A},\mbf{n}) = \det [\delta\mbf{\chi}/\delta\mbf{n}]$.
Next, the identity is inserted into the integral $\int D\mbf{A}_\mu
\exp(-S)$, with $S$ being the Yang-Mills action (gauge fixing
and Faddeev-Popov ghost terms are not written explicitly), then $\mbf{A}_\mu$
is replaced by $\mbf{n}C_\mu + \mbf{W}_\mu$, with a {\em generic}
 $\mbf{W}_\mu$
perpendicular to $\mbf{n}$ so that $D\mbf{A}_\mu \sim
DC_\mu D\mbf{W}_\mu$. Finally, one shifts the integration variables
$\mbf{W}_\mu \rightarrow \mbf{W}_\mu + \mbf{\alpha}_\mu$. As a result
one arrives at the following representation
\be
{\cal Z} \sim \int D\mbf{A}_\mu\, e^{-S}\sim 
\int D\mbf{n} DC_\mu D\mbf{W}_\mu \Delta(\mbf{A},\mbf{n})
\delta[\mbf{\chi}(\mbf{W},C,\mbf{n})]\, e^{-S}\ .
\label{t1}    
\ee
In the integrand of the right-hand side of Eq. (\ref{t1}), 
$\mbf{A}_\mu$ must be replaced
by (\ref{1}). The integral over $\mbf{W}_\mu$ seems to depend
on the choice of $\mbf{\chi}$. However, this is not always the case.

Various
choices of $\mbf{\chi}$ can be regarded as {\em gauge fixing}
conditions for the gauge symmetry associated with a reparameterization
of $\mbf{W}_\mu$. As it stands, Eq.(\ref{1})
contains 14 functions in the right-hand side, while there are only 12
components in $\mbf{A}_\mu$. Therefore the gauge transformations
(\ref{3}) would, in general, be induced by {\em five}-parametric
transformations of the new variables. There are two-parametric
transformations of the new variables under which $\mbf{A}_\mu$
remains invariant. Precisely this gauge freedom is fixed by (\ref{1a})
and by the corresponding delta function in (\ref{t1}). 
The key point is that the invariance of the integral over $\mbf{W}_\mu$ in
(\ref{t1}) under variations of $\mbf{\chi}$ can be established just as 
the gauge invariance of the perturbative 
Faddeev-Popov path integral is proved. Since
$\Delta$ is a determinant, it can be lifted up to the exponential
by introducing ghosts $\mbf{\eta}$
(which should not be 
confused with the conventional Faddeev-Popov ghosts), 
and $\delta(\mbf{\chi})$
is replaced by $\int D\mbf{f} \exp(-\mbf{f}^2/2)\delta(\mbf{\chi}-\mbf{f})$.
A change of
$\mbf{\chi}$ is equivalent to some BRST transformation
of $\mbf{\eta}$ and the new variables. When $\mbf{W}_\mu$ is
integrated out, the invariance of the effective action
for the remaining variables under variations of $\mbf{\chi}$
should be guaranteed by the invariance under the corresponding BRST
transformations of $\mbf{n}$ and $C_\mu$. Now we recall that
a change of $\mbf{\chi}$ implies a modification of the gauge
transformation law of $\mbf{n}$ and $C_\mu$. But for
all Abelian projections in question 
$\mbf{\chi}$ varies within the special class for which
$\mbf{n}$ transforms
according to (\ref{12a}), that is, 
neither the gauge transformation of 
$\mbf{n}$ nor $C_\mu$ depend on $\mbf{W}_\mu$.
Hence, the BRST transformations of $\mbf{n}$ and
$C_\mu$ generated by varying $\mbf{\chi}$ 
cannot be anything, but a subset of the conventional
BRST transformations associated with a gauge fixing in the original
integral over $\mbf{A}_\mu$. Owing to the 
BRST invariance of the Faddeev-Popov action,
we conclude that the effective action for $\mbf{n}$ and 
$C_\mu$ will also be invariant
under the BRST transformations generated
by variations of $\mbf{\chi}$. In short, one can say that
't Hooft's conjecture is a simple consequence of the gauge
invariance. In the new nonlocal variables, a relevance of the 
gauge symmetry is obvious, while in the original variables
it is less evident because of singularity of gauges used
in Abelian projections. The general case
when (\ref{12a}) is not valid will be considered elsewhere.  

If the dual scenario takes place, as suggested
by lattice simulations, the effective action for 
the topological current
$J_\mu$ has to be of the London type (as for superconductor).
Since the Abelian theory (\ref{9}) is SU(2) invariant,
the U(1) symmetry associated with the conservation of $J_\mu$ can be
dynamically broken {\em regardless} of any gauge fixing used
to compute the functional integral. By means of the representation
(\ref{t1}), where the integration over the field $\mbf{n}$ provides
the sum over topological configurations of Yang-Mills fields,
we have circumvented a difficult problem of summing over
monopole configurations in singular Abelian projection gauges.

{\bf 5. Partial duality}. 
The homotopy group arguments show that the field $\mbf{n}$
may also contain configurations that upon the 
Abelian projection $\mbf{n}\rightarrow \mbf{n}_0$ form 
closed magnetic fluxes which are linked and/or knotted.
Their topological number is known as the Hopf invariant and 
associated with the nontriviality of $\pi_3(S^2)$ of the map $\mbf{n}$.
Due to a nonlocality of the Hopf invariant, there is no conservative 
current related to such topological defects.
The magnetic fluxes cannot be observed in numerical studies
by the same procedure as that used for magnetic monopoles 
because they do not contribute to the total magnetic field flux
through any closed surface. It has been conjectured
that quantum fluctuations of other degrees of freedom of
Yang-Mills fields may stabilize  fluxes against shrinking so
that they would behave like knot solitons \cite{5}. The dynamical regime
in which fluxes exist as physical excitations is dual to some
Higgs phase which is revealed via a special parameterization 
of the Yang-Mills connection \cite{5}. 

To verify this conjecture, one needs a more general parameterization
of the connection than that used in \cite{5} in order to 
correctly describe an {\em off-shell} quantum dynamics 
of relevant physical degrees of freedom. The problem is
to find an {\em explicit} and {\em local} parameterization
of $\mbf{W}_\mu$ by six functional variables, while 
keeping the partial duality between $\mbf{n}$ and some
components of $\mbf{W}_\mu$. Then in the 
new variables the Yang-Mills theory will be a local
Abelian theory (\ref{9}) to which the Wilsonian arguments
of \cite{5} can be applied. 

The necessary six functional variables can be unified into
 an antisymmetric tensor $W_{\mu\nu}=-W_{\nu\mu}$.
Consider the following representation 
\be
\mbf{W}_\mu = g \left[W_{\mu\nu} + 
V_{\mu\nu}(W,\mbf{n})\right]\mbf{\alpha}_\nu
\ ,\label{f1}
\ee
where $V_{\mu\nu}$ is a symmetric tensor which depends on $W_{\mu\nu}$
and $\mbf{n}$.
This is the most general form of $\mbf{W}_\mu$. 
It should be noted that $W_{\mu\nu}$ is {\em dimensionless}
just as the topological field $\mbf{n}$, which is necessary 
for $W_{\mu\nu}$ to be a dual variable to $\mbf{n}$.
In principle,
one can take a generic isotopic vector $\mbf{\gamma}_\mu(\mbf{n})$,
perpendicular to $\mbf{n}$, instead of $\mbf{\alpha}_\mu$ in (\ref{f1}). 
However, by a redefinition of the 
symmetric and antisymmetric components of the tensor $W_{\mu\nu}+V_{\mu\nu}$,
$\mbf{\gamma}_\mu$ can always be replaced by $\mbf{\alpha}_\mu$ because
any isotopic vector perpendicular to $\mbf{n}$ is a linear
combination of the Lorentz components of $\mbf{\alpha}_\mu$. 
The simplest choice $V_{\mu\nu}=0$ would already provide us with
a sought-for parameterization to develop the off-shell dynamics
of the physical degrees of freedom. It implies only {\em one}
gauge condition on a generic connection (\ref{1}), while 
we are allowed to impose {\em three} without solving the Gauss law.
Indeed, if $V_{\mu\nu} =0$, $\mbf{W}_\mu$ satisfies {\em three}
(not {\em two} as required by (\ref{1a})) equations
\be
\mbf{W}_\mu\otimes\mbf{\alpha}_\mu + \mbf{\alpha}_\mu
\otimes\mbf{W}_\mu =0\ .
\label{f3}
\ee
The tensor product contains three independent components because
both $\mbf{W}_\mu$ and $\mbf{\alpha}_\mu$ are perpendicular
to $\mbf{n}$. Therefore there is one constraint on the components
of $W_{\mu\nu}$: $W_{\mu\nu}H_{\mu\nu}=0$. 
Since for the functional integral one needs a 
{\em change } of variables, the latter restriction 
on $W_{\mu\nu}$ can be relaxed
to achieve this goal if, for example, we set
\be
\mbf{W}_\mu = gW_{\mu\nu}\mbf{\alpha}_\nu +g\rho\mbf{\alpha}_\mu\ ,
\label{f4}
\ee
where $\rho=\rho(W,\mbf{n})\sim W_{\mu\nu}H_{\mu\nu}$ 
is determined by (\ref{f3}). The field $\rho$ in (\ref{f4}) is,
in general, specified modulo a factor which may depend on $\mbf{n}$.
For instance, the condition (\ref{f3}) can be modified by
multiplying each of the two terms in the tensor product
by coefficients depending on $\mbf{n}$.

Thanks to the gauge invariance of the Yang-Mills action, a
particular choice of $\rho$ should not be relevant for
the partial duality because $\rho$ can always be removed by
an appropriate  gauge transformation (\ref{5}). 
Quantum dynamics of the charged fields in the Abelian theory
(\ref{9}) is described by the antisymmetric field $W_{\mu\nu}$.
The Jacobian of the change of variables is the determinant
of the Euclidean metric $ ds^2= \int dx d\mbf{A}_\mu\cdot d\mbf{A}_\mu$ 
on the affine space of connections in the new variables 
(\ref{1}) and (\ref{f4}); $d\mbf{A}_\mu$ denotes a functional
differential of the affine (field) coordinate $\mbf{A}_\mu$. 
The Jacobian induces quantum corrections, 
associated with the curvilinearity
of the new field variables, to the classical action (\ref{9}).
If the dynamics
of the charged field $W_{\mu\nu}$ 
is such that the average over them yields
\be
\left\<\left(\pl_\mu W_{\nu\sigma}-\pl_\nu W_{\mu\sigma}\right)
\left(\pl_\mu W_{\nu\lambda}-\pl_\nu W_{\mu\lambda}\right)\right\>
\sim m^2 \delta_{\sigma\lambda}\ ,
\label{20}
\ee
then in the large distance limit, the leading term of the gradient
expansion of the effective action for the field $\mbf{n}$ would
contain the term $m^2\mbf{\alpha}_\mu\cdot\mbf{\alpha}_\mu 
= m^2(\pl_\mu\mbf{n})^2$.
Together with the tree level term proportional to $H_{\mu\nu}^2$,
it forms, as follows from (\ref{9}),
the action of the Faddeev model \cite{8} which
describes knot-like massive solitons. Such solitonic
excitations could be good candidates for glueballs.
Their stability in the effective theory depends on other terms
which are contained in the gradient
expansion of the effective action. 

Consider the decomposition $\pl_\mu \mbf{n} =b_\mu^*\mbf{n}_r
+b_\mu\mbf{n}_r^*$. We have $H_{\mu\nu}=i g^{-1}(b_\mu^*b_\nu
-b_\mu b^*_\nu)$ and $W_\mu = iW_{\mu\nu}b_\mu +i\rho b_\mu$.   
The dual (Higgs) phase reported in \cite{5}
may also exist in the Abelian theory (\ref{9}), provided the average
over the field $\mbf{n}$ has the property that 
\be
\<b_\mu b_\nu\>=0\ ,\ \ \ \ 
\<b_\mu^*b_\nu\> \sim M^2\delta_{\mu\nu}\ .
\label{f5a}
\ee
In particular, the property (\ref{f5a}) implies that $\<H_{\mu\nu}\>=0$
and $\<H_{\mu\sigma}H_{\nu\lambda}\> \approx 2g^{-2}
M^4(\delta_{\mu\nu}\delta_{\sigma\lambda}-\delta_{\mu\lambda}
\delta_{\sigma\nu})$ (neglecting by a four-point function
of the field $b_\mu$). 
The effective potential
for $W_{\mu\nu}$ would have  ``classical'' minima:
$\<W_{\mu\sigma}W_{\nu\sigma}\>$ $ \sim\delta_{\mu\nu}$, therefore the
Maxwell field acquires a mass proportional to $M$.

The parameterization relevant for the partial duality is given
by the first term in (\ref{f1}). Therefore the choice of $V_{\mu\nu}$
does not seem to be important. This suggests that the property
(\ref{20}) should be universal relative to a choice of $V_{\mu\nu}$.
The Ansatz (\ref{f1}) can be used to solve Eq. (\ref{1a}) for $V_{\mu\nu}$.
In this case $V_{\mu\nu}$ may even be nonlocal (cf., e.g., (\ref{12b})).
It would be interesting to find arguments to prove that (\ref{20}) holds 
for any $V_{\mu\nu}=V_{\mu\nu}(W,\mbf{n})$ if it holds for at least
one choice of $V_{\mu\nu}$. This amounts to the existence of the gauge
$V_{\mu\nu}=0$ for any $\mbf{\chi}$ (for algebraic conditions like (\ref{f3}),
this is the case). The nontriviality of the problem
is that the gauge transformation law of the new variables depends 
on $\mbf{\chi}$.

{\bf 6. Gauge group SU(N)}. To extend our description of the Yang-Mills
theory as an Abelian theory with topological degrees of freedom
to the gauge group SU(N),
we introduce the Cartan-Weyl basis in the Lie algebra \cite{6}. Let
$\omega_k$ be simple roots, $k=1,2,...,N-1$ (= rank of SU(N)), and
$\beta$ be a positive root. It can be written in the form
$\beta = \omega_{k}+\omega_{k+1}+\cdots + \omega_{k+j}$.
All simple roots have the same norm. The angle between 
$\omega_k$ and $\omega_{k\pm 1}$ is $2\pi/3$, while
$\omega_k$ and $\omega_{k\pm j}$, $j\geq 2$, are perpendicular.
As a consequence, all roots have the same norm.
For every root
$\beta$ two basis elements $e_\beta$ and $e_{-\beta}=e^*_\beta$ 
are defined so that 
\be
[h,e_\beta]= (h,\beta)e_\beta\ ,\ \ \ \ 
[e_\beta,e_\gamma]= N_{\beta,\gamma}e_{\beta+\gamma}\ , \ \ \ \
[e_\beta,e_{-\beta}]= \beta \ ,
\label{11}
\ee
where $h$ is  any element from the Cartan subalgebra; and
for any two elements $v$ and $w$ of the Lie algebra
the Killing form is defined by $(v,w)={\rm tr}\, ({\rm ad}\,v
{\rm ad}\,w)$. The operator ${\rm ad}\,v$ acts on
any element $w$ as $[v,w]$. The structure constants $N_{\beta,\gamma}=
-N_{-\beta,-\gamma}$ are not zero only if $\beta +\gamma$ is a root.
For SU(N), $N_{\beta,\gamma}^2 = (2N)^{-1}$ and relative signs
can be fixed by the Jacobi identity for the basis elements. 
Let $h_k=h_k^*$ be an orthonormal basis with respect to the 
Killing form in the Cartan subalgebra. With the normalization 
of the structure constants as given in (\ref{11}), the elements
$h_k$, $e_\beta$ and $e^*_\beta$ form an orthonormal basis in
the Lie algebra, $(h_k,e_\beta)=0$, 
$(e_\beta,e_\gamma)=0$ and
$(e_\beta,e_\gamma^*)=\delta_{\beta\gamma}$.
 
Let $U=U(x)$ be a generic element of the 
coset $SU(N)/[U(1)]^{N-1}$. Consider a local orthonormal basis
$n_k = U^\dagger h_k U$, $n_\beta = U^\dagger e_\beta U$.
The commutation relations
(\ref{11}) hold for the local basis too. For any element $v$
one can prove the identity
\be
v = N[n_k,[n_k,v]] + n_k(n_k,v)\ .
\label{12}
\ee
A proof is based on a straightforward computation of the 
double commutator in (\ref{12}) in the Cartan-Weyl basis
and the fact that all roots of SU(N) have the same norm
which is $(\beta,\beta) = 1/N$ relative to the Killing form.
A {\em change} of variables  
in the affine space of SU(N) connections reads
\be
A_\mu = \alpha_\mu + n_k C^k_\mu + W_{\mu}\ ,\ \ \ 
\alpha_\mu = ig^{-1}N[\pl_\mu n_k,n_k]\ ,\ \ \
(W_\mu,n_k)=0\ ,
\label{13c}
\ee
where $W_\mu$ is subject to $N^2-N$ conditions
$\chi(W,C^k,n_k)=0$, $(\chi,n_k)\equiv 0$
Thus, in  four dimensional spacetime,  
$4(N^2-1)$ independent components of $A_\mu$ are now
represented by $N^2-N=\dim SU(N)/[U(1)^{N-1}]$ independent components
of $n_k$, by $4(N-1)$ components of $C_\mu^k$ and by $3(N^2-N)$
components of $W_{\mu}$.
The two first terms in (\ref{13c}) are constructed so that the 
corresponding field strength is purely Abelian in the local
basis
\be
F_{\mu\nu}(\alpha +C)= n_k\left(C^k_{\mu\nu} - H^k_{\mu\nu}\right)
\equiv n_k F^k_{\mu\nu}\ ,\ \ \ 
H_{\mu\nu}^k = ig^{-1}N \left(n_k,[\pl_\mu n_j,\pl_\nu n_j]\right)\ ,
\label{14}
\ee
and $C_{\mu\nu}^k = \pl_\mu C^k_\nu -\pl_\nu C^k_\mu$. 
Relation (\ref{14}) is obtained from the definition
$F_{\mu\nu}(A)=\pl_\mu A_\nu -\pl_\nu A_\mu +ig[A_\mu,A_\nu]$
by a successive use of the Jacobi identity and (\ref{12}).

The homotopy groups $\pi_2(G/H)$ and $\pi_3(G/H)$, where
$G=SU(N)$ and $H=[U(1)]^{N-1}$, of the map $n_k$ are nontrivial.
Therefore $n_k$ carry topological (physical) degrees of freedom
of Yang-Mills fields.  
It is not hard to establish the identity
\be
ig^{-1}\pl_\mu U^\dagger U = \alpha_\mu + n_k H^k_\mu\ ,\ \ \ \
H_{\mu\nu}^k = \pl_\mu H^k_\nu - \pl_\nu H^k_\mu - H_{\mu\nu}^{k(st)}\ ,
\label{16}
\ee
where the group element $U$
specifies an orientation of the local basis with respect to the Cartan-Weyl
basis, and $H_{\mu\nu}^{k(st)} = ig^{-1}(n_k,[\pl_\mu,\pl_\nu]U^\dagger U)$
is the field of Dirac strings. The Abelian strength tensor 
$F_{\mu\nu}^k$ does not satisfy the Bianchi identity if the 
monopole-like defects 
associated with a nontriviality of $\pi_2(G/H)$ are present.
The conservative topological current is a simple multi-component
generalization of (\ref{11b}): $J_\mu^k =\pl_\nu\,{}^\star\! F_{\mu\nu}^k$,
$\pl_\mu J_\mu^k=0$. There is a (magnetic) symmetry group $[U(1)]^{N-1}$
responsible for its conservation.

To reformulate the Yang-Mills theory as an Abelian theory
without gauge fixing, we introduce the decomposition
 $W_\mu = W_\mu^{\beta *}n_\beta + W_\mu^\beta n^*_\beta$
and set $A_\mu^k = C_\mu^k - H_\mu^k$.
The Lagrangian density in the new variables assumes the form
\ba
F_{\mu\nu}^2& =& \left\{C_{\mu\nu}-H_{\mu\nu}
+ig(h_k,\beta)\left[
W_\mu^{\beta *}W_\nu^\beta -W_\nu^{\beta *}W_\mu^\beta\right]\right\}^2
\nonumber\\
&\ &
+ \left\vert D_\mu W_\nu^\beta - D_\nu W^\beta_\mu + 
ig \Gamma_{\mu\nu}^\beta
\right\vert^2 \ , \label{18a}\\
\Gamma_{\mu\nu}^\beta&=& \sum_{\alpha 
+\gamma=\beta}N_{\alpha,\gamma}\left[
W_\mu^{\alpha }W_\nu^{\gamma }-W_\nu^{\alpha }W_\mu^{\gamma }\right]
+ \sum_{\alpha -\gamma=\beta}N_{\alpha,-\gamma}\left[
W_\mu^{\alpha }W_\nu^{\gamma *} -W_\nu^{\alpha }W_\mu^{\gamma *}\right]\ ,
\label{18b}
\ea
where $\alpha,\beta$ and $\gamma$ are positive roots, and
$D_\mu W_\nu^\beta = \pl_\mu W_\nu^\beta -ig(\beta,h_k)A_\mu^kW_\nu^\beta$.
A calculation of the field strength tensor is somewhat tedious
but straightforward. The identities $\nabla_\mu(\alpha)n_\beta
\equiv \pl_\mu n_\beta +ig[\alpha_\mu,n_\beta]=-
ig (h_k,\beta)H_k n_\beta$ and
$(\pl_\mu n_j,n_k)
= (\pl_\mu n_k,n_j)=0$, which can be deduced from (\ref{16}),
are useful to simplify the computation.
The Lagrangian density (\ref{18a}) is invariant under the Abelian
gauge transformations
\be
A_\mu^k\rightarrow A_\mu^k + \pl_\mu\xi_k\ ,\ \ \ \
W_\mu^\beta \rightarrow e^{ig(\beta,h_k)\xi_k}W_\mu^\beta\ .
\label{19}
\ee
The Abelian gauge group $[U(1)]^{N-1}$
is not a subgroup of the original gauge
group. Just as in the SU(2) case, it is related to the fact that
the basis elements $n_\beta$ can be transformed locally
\be
n_\beta \rightarrow e^{i(\beta,h_k)\xi_k}n_\beta\ 
\label{19a}
\ee
without spoiling both the orthogonality and commutation relations 
in the local Cartan-Weyl basis.

The gauge transformation law of the new variables can be found
by the same method as in the SU(2) case. It depends on the choice
of $\chi$. There is a wide class of $\chi$'s, 
associated with Abelian projection gauges as explained in section 3,
for which the gauge transformation law has a particularly simple form
\be
\delta n_k = i[n_k,\varphi]\ ,\ \ \ \ 
\delta C_\mu^k = g^{-1} (n_k,\pl_\mu \varphi)\ ,\ \ \ \
\delta W_{\mu} 
= i[W_{\mu}, \varphi]
\ . \label{15}
\ee
In this case,
the SU(N) gauge transformations are generated by  
adjoint transformations of any {\em rigid} local
Cartan-Weyl basis, under which
the field variables in the Abelian theory are invariant
\be
\delta A^k_\mu =0\ ,\ \ \ \   \delta W_\mu^\beta = 0\ .
\label{17}
\ee
All Abelian projections in which topological defects occur 
in Abelian components of projected connections fall into
one class defined by the projection $n_k\rightarrow h_k$
in the new variables. A proof of
the 't Hooft conjecture is a straightforward generalization 
of the SU(2) case. The key point is the gauge symmetry
(\ref{15}) which does not mix $W_\mu$ with the Abelian
variables $C_\mu^k$ and $n_k$. Therefore the monopole
dynamics in the SU(N) Yang-Mills theory is projection independent.

The coupling constants of the interaction of $W_\mu^\beta$
among each other and with the Maxwell fields $A_\mu^k$
in (\ref{18a}) are proportional to $N^{-1/2}$ because
$N_{\beta,\gamma}\sim N^{-1/2}$ and $ \vert (\beta,h_k)\vert
\leq N^{-1/2}$, while $H_{\mu\nu}^k \sim N^{1/2}$.
Therefore in the large $N$ limit the dynamics of the 
topological fields $n_k$ dominates \cite{7}.

Finally, we observe that precisely in {\em four} dimensional
spacetime, the $3(N^2-N)$ independent components of $W_\mu$ 
can be unified into a tensor $W_{\mu\nu}^{jk}$ which
is {\em antisymmetric} in the Lorentz indices and {\em symmetric}
in the Cartan indices, 
$W_{\mu\nu}^{jk}=-W_{\nu\mu}^{jk}$
and $W_{\mu\nu}^{jk}=W_{\mu\nu}^{kj}$.
This suggests the following local parameterization to reveal
a partial duality in the SU(N) Yang-Mills theory
\be
W_\mu = \left\{ W_{\mu\nu}^{jk} + V_{\mu\nu}^{jk}(W,n)\right\}
\alpha^{jk}_\nu\ , \ \ \ \ 
\alpha_\mu^{jk}= i[\pl_\mu n_j,n_k]= \alpha_\mu^{kj}\ , 
\label{17a}
\ee
where the symmetric tensor $V_{\mu\nu}^{jk}$ can be specified
as an explicit and local function of $W_{\mu\nu}^{jk}$ and
$n_k$ via a simple generalization of the method of section 5.

{\bf 7. Conclusions}. 
An Abelian structure and the Abelian magnetic symmetry can be 
established in the SU(N) Yang-Mills theory without any
gauge fixing (or Abelian projections). By making use of 
such an Abelian theory we have shown
that the effective dynamics of the topological
degrees of freedom that are singled out as magnetic
monopoles in Abelian projections is independent of the
projection ('t Hooft conjecture). We have also generalized
a parameterization of Faddeev and Niemi in order to study
an off-shell dynamics of physical degrees of freedom
which may form knot-like solitons in the infrared 
region of the Yang-Mills theory. 
A general parameterization of the Yang-Mills connection
to reveal a partial duality between the topological field
$\mbf{n}$ and a dimensionless antisymmetric field 
$W_{\mu\nu}$ has been proposed. It is believed that
the separation of the topological degrees of freedom
of the Yang-Mills theory via a {\em change} of variables
in the functional integral is important for developing 
the corresponding effective action by analytical means.

\end{document}